\documentclass[pra,aps,onecolumn,showpacs]{revtex4}
\usepackage{amsmath,graphicx,epsfig}
\def\fig_width{3. in} 
\draft

\newlength{\defbaselineskip}
\newcommand{\setlinespacing}[1]%
           {\setlength{\baselineskip}{#1 \defbaselineskip}}

\begin{document}

\title{Erratum: Light-induced desorption of alkali-metal atoms from paraffin coating ~~~ [Phys. Rev. A 66, 042903 (2002)]} 
\author{E. B. Alexandrov}
\author{M. V. Balabas}
\affiliation{S. I. Vavilov State Optical Institute, St. Petersburg, 199034 Russia}
\author{D. Budker}
\email{budker@socrates.berkeley.edu} \affiliation{Department of Physics, University of
California at Berkeley, Berkeley, California 94720-7300} \affiliation{Nuclear Science
Division, Lawrence Berkeley National Laboratory, Berkeley, California 94720}
\author{D. English}
\author{D. F. Kimball}
\author{C.-H. Li}
\author{V. V. Yashchuk}
\affiliation{Department of Physics, University of California at Berkeley, Berkeley,
California 94720-7300}


\date{\today}



\pacs{PACS. 34.50.Dy, 79.20.La}




\maketitle

In the above manuscript, there was an error in the numeric computation [based on Eqs.~(6)
and (7)] predicting the time-dependent alkali vapor density $n(t)$ and number of alkali
atoms in the paraffin coating $N_c(t)$.  Once the error in the computation was remedied,
it was necessary to adjust two parameters [$\rho$ (probability for adsorption of atoms by
the paraffin coating per wall collision) and $\xi$ (exchange rate between atoms in the
stem of the cell and the volume of the cell)] in order to obtain the same agreement
between the theory and the data as seen in Figs.~9 and 10.  This modifies the results for
$\rho$ and $\xi$ listed in Table II as follows:

\addtocounter{table}{1}

\begin{table}[h]
\caption{Average values of various light-independent parameters from the model extracted
from fits to the data from Secs.~III B, III C, and III D.}
\medskip \begin{tabular}{ c c c }
\hline \hline
            ~~~~~Parameter~~~~~   & ~~~~~~~Type (a) cell~~~~~~~                         & ~~~~~~~Type (b) cell~~~~~~~                     \\
\hline
            $\Gamma$    & $5(2)\times10^{-4}~{\rm s^{-1}}$      & $7(4)\times10^{-4}~{\rm s^{-1}}$  \\
            $\gamma_d$  & $2(1)\times10^{-4}~{\rm s^{-1}}$      & $2(1)\times10^{-4}~{\rm s^{-1}}$  \\
            $\rho$      & $4.9(3)\times10^{-6}$                 & $4(2)\times10^{-6}$             \\
            $\xi$       & $5.0(2)~{\rm cm^3/s}$                 & $1.49(3)~{\rm cm^3/s}$           \\
\hline \hline
\end{tabular}
\label{FittedParameterTable}
\end{table}

Therefore the average over all data for $\rho$ should be $4(1) \times 10^{-6}$.  This in
fact improves agreement with the values of $\rho$ determined in other work
\cite{Bouchiat,Liberman,Balabas,Balabas2}, in which $\rho$ was found to be between $\sim
10^{-6}$ and $\sim 3 \times 10^{-6}$.  This in turn means that the number of atoms
estimated to be in the coating $N_c$ should be larger by a factor of about $30$ [see, for
example, Eq.~(9)].  This would correspondingly change the scale of the lower plots in
Figs. 9 and 10.

The adjustment to the value for $\xi$ changes our estimate of the number of atoms
adsorbed by the coating during the ``ripening'' process to $\sim 10^{17}$, but this in no
way changes our conclusion that the number of atoms available for desorption is
considerably smaller than the number of atoms absorbed during cell ripening.

\bigskip

We are deeply indebted to A. I. Okunevich for bringing this mistake to our attention.

\end{document}